\documentclass[aps,prl,twocolumn,superscriptaddress,showpacs]{revtex4-1}   % appearance in journal
\usepackage{lineno}
\usepackage{amsmath}
\usepackage{amsfonts}
\usepackage{multirow}
\usepackage{verbatim}
\usepackage{color}     
\usepackage{array}
\usepackage{dcolumn}
\usepackage{xspace}
\usepackage{textcomp}
\usepackage{graphicx}
\usepackage{gensymb}
\usepackage{textgreek}
\usepackage[colorlinks,linkcolor=blue,citecolor=blue,urlcolor=blue]{hyperref}

\newcommand{\Duke}{%
	Department of Physics, Duke University, Durham, North Carolina 27708-0308, USA}
\newcommand{\TUNL}{%
	Triangle Universities Nuclear Laboratory, Durham, North Carolina 27708-0308, USA}
\newcommand{\NCCU}{%
	Department of Mathematics and Physics, North Carolina Central University, Durham, North Carolina, 27707, USA}
\newcommand{\GWUINT}{%
	Institute for Nuclear Studies, Department of Physics, The George Washington University, Washington DC 20052, USA}
\newcommand{\UNC}{%
	University of North Carolina at Chapel Hill, Chapel Hill, North Carolina 27516, USA}
\newcommand{\UKY}{%
	Department of Physics and Astronomy, University of Kentucky, Lexington, Kentucky 40506, USA}
\newcommand{\MSU}{%
	Department of Physics and Astronomy, Montclair State University, Montclair, New Jersey 07043, USA}	
\newcommand{\USask}{%
	Department of Physics and Engineering Physics, University of Saskatchewan, Saskatoon, Saskatchewan, S7N 5E2, Canada}
\newcommand{\NGC}{%
	Department of Physics and Astronomy, University of North Georgia, Dahlonega, Georgia 30597, USA}
\newcommand{\JMU}{%
	Department of Physics and Astronomy, James Madison University, Harrisonburg, Virginia 22807, USA}

\newcommand{\UM}{%
    Theoretical Physics Group, School of Physics and Astronomy, University of Manchester, Manchester M13 9PL, UK}

\begin{document}
 
\title{Proton Compton Scattering from Linearly Polarized Gamma Rays}

\author{X.~Li}
\email[]{xqli@mit.edu}
\affiliation{\Duke{}}
\affiliation{\TUNL{}}

\author{M.W.~Ahmed}
\affiliation{\TUNL{}}
\affiliation{\NCCU{}}

\author{A.~Banu}
\affiliation{\JMU{}}

\author{C.~Bartram}
\affiliation{\TUNL{}}
\affiliation{\UNC{}}

\author{B.~Crowe}
\affiliation{\TUNL{}}
\affiliation{\NCCU{}}

\author{E.J.~Downie}
\affiliation{\GWUINT{}}

\author{M.~Emamian}
\affiliation{\TUNL{}}

\author{G.~Feldman}
\affiliation{\GWUINT{}}

\author{H.~Gao}
\affiliation{\Duke{}}
\affiliation{\TUNL{}}

\author{D.~Godagama}
\affiliation{\UKY{}}

\author{H.W.~Grie{\ss}hammer}
\affiliation{\GWUINT{}}
\affiliation{\Duke{}}

\author{C.R.~Howell}
\affiliation{\Duke{}}
\affiliation{\TUNL{}}

\author{H.J.~Karwowski}
\affiliation{\TUNL{}}
\affiliation{\UNC{}}

\author{D.P.~Kendellen}
\affiliation{\Duke{}}
\affiliation{\TUNL{}}

\author{M.A.~Kovash}
\affiliation{\UKY{}}

\author{K.K.H.~Leung}
\affiliation{\Duke{}}
\affiliation{\TUNL{}}
\affiliation{\MSU{}}

\author{D.M.~Markoff}
\affiliation{\TUNL{}}
\affiliation{\NCCU{}}

\author{J.A.~McGovern}
\affiliation{\UM}

\author{S.~Mikhailov}
\affiliation{\TUNL{}}

\author{R.E.~Pywell}
\affiliation{\USask{}}

\author{M.H.~Sikora}
\affiliation{\GWUINT{}}
\affiliation{\TUNL{}}

\author{J.A.~Silano}
\affiliation{\TUNL{}}
\affiliation{\UNC{}}

\author{R.S.~Sosa}
\affiliation{\NCCU{}}

\author{M.C.~Spraker}
\affiliation{\NGC{}}

\author{G.~Swift}
\affiliation{\TUNL{}}

\author{P.~Wallace}
\affiliation{\TUNL{}}

\author{H.R.~Weller}
\affiliation{\Duke{}}
\affiliation{\TUNL{}}

\author{C.S.~Whisnant}
\affiliation{\JMU{}}

\author{Y.K.~Wu}
\affiliation{\Duke{}}
\affiliation{\TUNL{}}

\author{Z.W.~Zhao}
\affiliation{\Duke{}}
\affiliation{\TUNL{}}

\begin{abstract}
Differential cross sections for Compton scattering from the proton have been measured at scattering angles of $55\degree$, $90\degree$, and $125\degree$ in the laboratory frame using quasimonoenergetic linearly (circularly) polarized photon beams with a weighted mean energy value of 83.4\,MeV (81.3\,MeV). These measurements were performed at the High Intensity Gamma-Ray Source facility at the Triangle Universities Nuclear Laboratory. The results are compared to previous measurements and are interpreted in the chiral effective field theory framework to extract the electromagnetic dipole polarizabilities of the proton, which gives $\alpha_{E1}^p = 13.8\pm1.2_{\rm stat}\pm0.1_{\rm BSR}\pm0.3_{\rm theo},
\beta_{M1}^p = 0.2\mp1.2_{\rm stat}\pm0.1_{\rm BSR}\mp0.3_{\rm theo}$ in units of 10$^{-4}$\, fm$^3$.
\end{abstract}

\maketitle

The static electric and magnetic dipole polarizabilities of the proton $\alpha_{E1}^p$ and $\beta_{M1}^p$, respectively, reveal the internal dynamics of the proton. They parametrize the response of the proton's internal degrees of freedom to an external electromagnetic field. Considerable efforts have been taken to study the proton polarizabilities both experimentally and theoretically~\cite{Schumacher:2005an,Griesshammer:2012we,Hagelstein:2015egb}. In addition, $\beta_{M1}^p$ has been shown to be a crucial input in the determination of the two-photon-exchange contribution to the Lamb shift in muonic hydrogen~\cite{Birse:2012eb}. For the nucleon in general, the isovector difference $\beta_p-\beta_n$ has been connected to the nucleon electromagnetic mass difference, most recently in Refs.~\cite{Gasser:2015dwa, Walker-Loud:2019qhh}. The calculation of nucleon polarizabilities is also an aim of lattice QCD, and several groups now have published results, albeit almost all at large pion masses; see Ref.~\cite{Detmold:2019ghl} and references therein. 

Compton scattering has proven to be a powerful tool to probe the proton polarizabilities. At incident photon energies far below the pion-production threshold, Compton scattering from the proton can be described as the elastic scattering of the photon from a pointlike charged particle with an anomalous magnetic moment. As the photon energy increases, the effect of the electromagnetic polarizabilities as the leading-order structure-dependent contribution to the Compton scattering cross section becomes more significant. Compton scattering experiments below the pion-production threshold using liquid hydrogen targets and tagged or untagged bremsstrahlung photon beams have been performed at several gamma-ray source facilities in the last three decades. The Compton scattering cross sections on the proton have been extracted with significantly improved precision. To determine $\alpha_{E1}^p$ and $\beta_{M1}^p$ from Compton scattering, rigorous theoretical calculations are needed to fully describe the process. Chiral effective field theory ($\chi$EFT) has proven to be successful in interpreting Compton scattering data in terms of low-energy degrees of freedom~\cite{Griesshammer:2012we}. Currently the best determination of the Baldin sum rule (BSR) provides the constraint~\cite{Gryniuk:2015eza} 
\begin{equation}    
\alpha_{E1}^p+\beta_{M1}^p=14.0\pm0.2,
\label{eq:BSR}
\end{equation}
where the polarizabilities are given here and throughout this Letter in units of 10$^{-4}$\, fm$^3$. Applying this sum-rule constraint, the latest $\chi$EFT fit to the global database of proton Compton scattering below 200 MeV gives~\cite{McGovern:2012ew,Griesshammer:2015ahu}
\begin{equation}
\begin{split}
\alpha_{E1}^p &= 10.75\pm0.35_{\rm stat}\pm0.1_{\rm BSR}\pm0.3_{\rm theo},\\
\beta_{M1}^p &= 3.25\mp0.35_{\rm stat}\pm0.1_{\rm BSR}\mp0.3_{\rm theo}.
\end{split}
\label{eq:EFTproton}
\end{equation}
(This is an update of the published result which used a slightly different BSR value~\cite{deLeon:2001dnx}.)

The advent of the free-electron-laser (FEL)-based High Intensity Gamma-ray Source (HIGS) opens up new opportunities for nuclear Compton scattering~\cite{Griesshammer:2012we,Hagelstein:2015egb,Weller:2009zza,Myers:2012xw,Myers:2014qaa,Sikora:2017rfk,Li:2019irp}. This Letter reports the first proton Compton scattering experiment at HIGS using polarized photons, and the first analysis of absolute differential cross sections of proton Compton scattering using linearly polarized gamma-ray beams below the pion-production threshold. A $\chi$EFT fit has been performed on the Compton scattering cross sections on the proton obtained in the present experiment to extract the proton polarizabilities. The results are compared to the previously extracted values. The influence of this new extraction of the proton polarizabilities and the potential of further high-precision measurements are also discussed.

The experiment was performed at the HIGS facility at the Triangle Universities Nuclear Laboratory~\cite{Weller:2009zza}. The HIGS facility utilizes a FEL to produce intense, quasimonoenergetic, and nearly 100\% polarized gamma-ray beams via Compton backscattering~\cite{Yan:2019bru,Wu:2015hta,Wu:2006zzc}. 
The gamma-ray beam pulses are about 10\,ns wide separated by 179\,ns, allowing for clear identification of beam-related events and rejection of backgrounds in the detection system. The gamma-ray beam was collimated by a 25.4-mm-diameter lead collimator. 
The collimated beam flux was measured and continuously monitored using a collection of plastic scintillators with an internal radiator~\cite{Pywell09} by detecting the charged particles produced in the radiator. 
In this experiment, the gamma-ray beams were linearly (circularly) polarized at an intensity-weighted average energy of 83.4 (81.3)\,MeV with an energy spread of 2.7\% (6.5\%) FWHM. The on-target intensity of the photon beam was $\approx10^7\,\gamma/\rm{s}$. 

The photons were scattered from a cryogenic target of liquid hydrogen contained in a 20-cm-long conical frustum cell~\cite{Kendellen16}. The wall and end caps of the cell were made from 0.125-mm-thick Kapton foil. Scattering data were taken with the target cell both full of liquid hydrogen with an areal target density of $(8.40\pm0.08)\times10^{23}\,\rm{nuclei}/cm^{2}$ and empty in order to subtract the empty-target background. The scattered photons were detected using eight NaI(Tl) detectors (both in plane and out of plane) located at scattering angles of $\theta =$ 55$\degree$, 90$\degree$, and 125$\degree$ in the laboratory frame. Each detector consisted of a 25.4-cm-diameter, about 30-cm-long cylindrical core NaI(Tl) crystal surrounded by 7.5-cm-thick NaI(Tl) annular shield segments. The anticoincidence shield was used to veto cosmic-ray events. The acceptance cone of each detector was defined by a 15-cm-thick lead collimator installed in the front face of the NaI(Tl) detector. Five in-plane detectors (one at $\theta = 55\degree$, two at $\theta = 90\degree$, the other two at $\theta = 125\degree$) were placed on the tables with their axes and the linear polarization axis of the photon beam aligned in the same horizontal plane. The other three out-of-plane detectors (at $\theta =$ 55$\degree$, 90$\degree$, and 125$\degree$, respectively) were placed beneath this horizontal plane and pointed upward toward the target, with their axes aligned in a vertical plane perpendicular to the gamma-ray beam linear polarization direction. The geometry of the experimental apparatus (see more details in Refs.~\cite{Li:2019irp, Li:2020cvm}) was surveyed to a precision of 0.5\,mm and incorporated into a \textsc{geant4}~\cite{Agostinelli:2002hh} simulation to determine the effective solid angles of the detectors. The effective solid angle accounts for the geometric effects due to the extended target and the finite acceptance of the detectors, as well as the attenuation of scattered photons in the target cell and the surrounding materials. With the distance from the front face of each detector collimator aperture to the target center measuring about 58\,cm, the simulated effective solid angles ranged from 62.7 to 65.6\,msr.

\begin{figure*}[!hbt]
\centering
\includegraphics[width=0.99\textwidth]{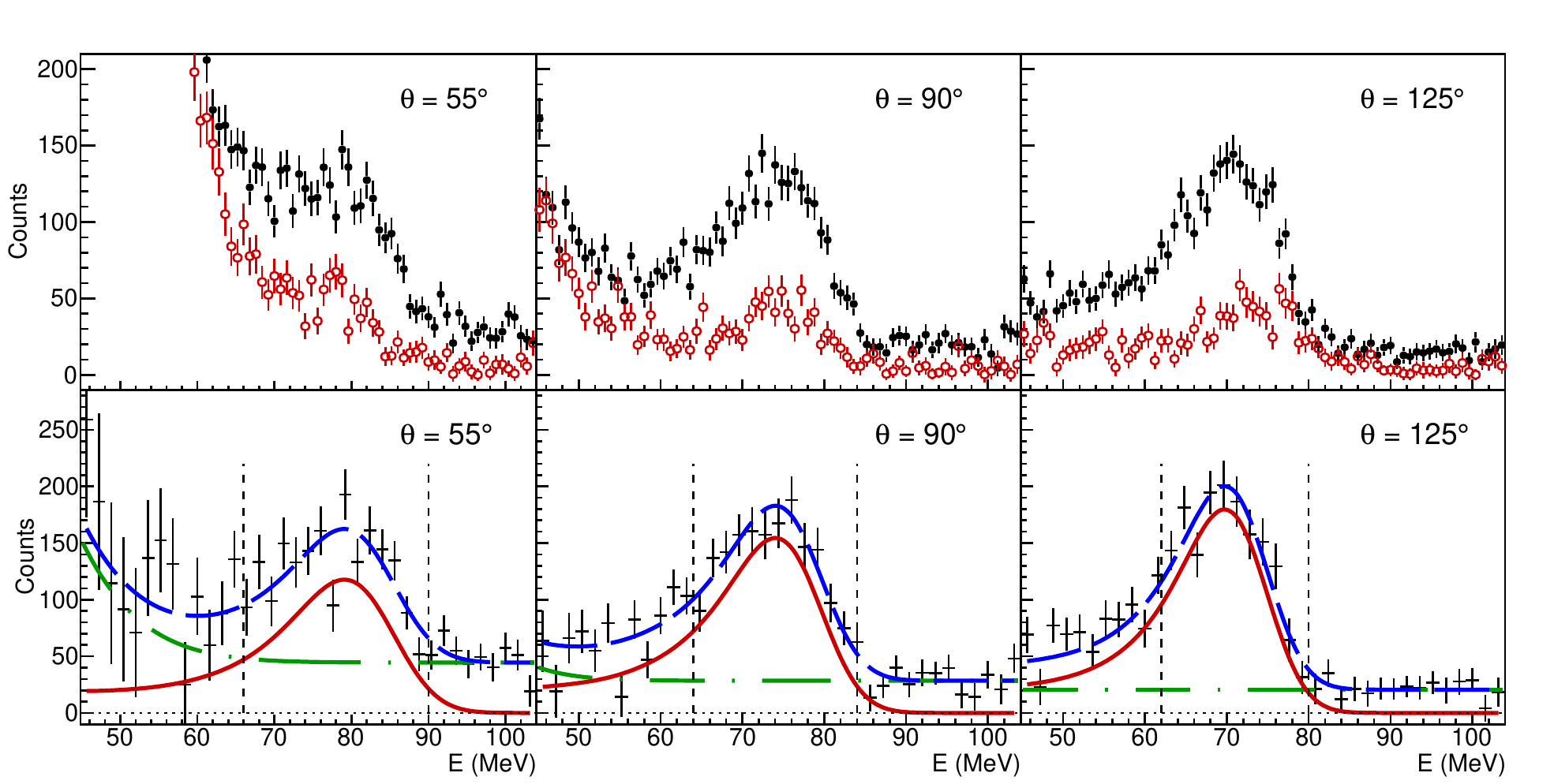}
\caption{Upper: representative energy spectra for full (closed circles) and empty (open circles) targets at 55$\degree$, 90$\degree$, and 125$\degree$. Lower: final energy spectra with empty-target events removed. The total fit (dashed curve) is the sum of the electromagnetic background (dot-dashed curve) and the \textsc{geant4} simulated detector response function (solid curve). The vertical dashed lines indicate the ROI. }
\label{fig:Energy_spectra}
\end{figure*}

The signals from the core and shield NaI detectors were recorded using a 14-bit digitizer with a sampling rate of 500\,MHz. A cut was applied on the energy spectrum from each shield detector to reject cosmic-ray events while losing a negligible number of Compton events. In addition, the pulsed nature of the photon beam produced a clear prompt timing peak for beam-related events atop a flat random background, allowing for a timing cut to veto a large portion of beam-unrelated backgrounds. The shield-energy cut and the timing cut on the prompt peak together removed over 99\% of the cosmic-ray events within the region of interest (ROI) in the energy spectrum. After applying both cuts, the remaining background from random, time-uncorrelated events was removed by sampling the energy spectrum in the random region and subtracting it in the prompt region after normalizing to the relative widths of the timing windows. The above analysis was applied to the full-target and empty-target data and the representative resulting energy spectra are shown in Fig.~\ref{fig:Energy_spectra}. For each detector, the final energy spectrum was obtained by subtracting the energy spectrum of the empty target from that of the full target after scaling to the number of incident photons. Typical final energy spectra at the three scattering angles are shown in Fig.~\ref{fig:Energy_spectra}.

The aforementioned \textsc{geant4} simulation of the full experimental apparatus was used to determine the line shape of the detector. Scattered photons were propagated from within the target volume to the detectors to simulate for the detector response functions. The energy of the simulated scattered photon $E_\gamma^{\prime}$ was
\begin{equation}
     E_\gamma^{\prime} = \frac{E_\gamma}{1 + \frac{E_\gamma}{Mc^{2}}(1-\cos\theta)},
\end{equation}
where $E_\gamma$ is the incident photon energy sampled from the beam energy profile taking into account the energy spread of the photon beam, $\theta$ is the laboratory scattering angle, $M$ is the proton mass, and $c$ is the speed of light. The simulated detector response was then convoluted with a Gaussian smearing function, which accounted for the intrinsic detector resolution, to obtain the elastic-scattering line shape. As reported in Refs.~\cite{Myers:2012xw,Myers:2014qaa}, the in-beam test using the same set of NaI(Tl) detectors has proven that the measured response function can be accurately reproduced with such a method. The smeared detector response and the backgrounds were simultaneously fitted to the final energy spectrum. For the forward-angle detectors, the background from atomic processes was prominent in the low-energy region and was fitted with an exponential function. Additional flat backgrounds from the scattering of the bremsstrahlung photons induced by the electron beam in the storage ring were fitted for all detectors. 
The number of events were summed from the background-subtracted spectrum over the ROI and then corrected by an efficiency factor to extract the photon yield. The efficiency factor was defined as the ratio of the integral of the line shape within the ROI to that of the entire energy range from 0 to 100\,MeV. To determine the differential cross section, the yield was normalized to the target thickness, the number of incident photons corrected for the average loss due to target absorption, the bin center correction factors, and the effective solid angles. 

The systematic uncertainties were grouped into two categories, point to point and overall normalization. The point-to-point uncertainties resulted from placing cuts in the timing and shield-energy spectra and locating boundaries of the ROI and the fitting window, which varied between individual detectors. Additional point-to-point systematic uncertainties were assigned to the two out-of-plane detectors at the scattering angles of $55^\circ$ and $125^\circ$ due to the insufficient measurement of the distances between them and the target. All contributions were summed in quadrature to obtain the total point-to-point systematic uncertainty for each data point ranging from 4.5\% to 13.8\%. The normalization uncertainties included the uncertainties from the number of incident photons (2\%) and the target thickness (1\%). 

\begin{figure}[!htp]
\centering
\includegraphics[width=1\columnwidth]{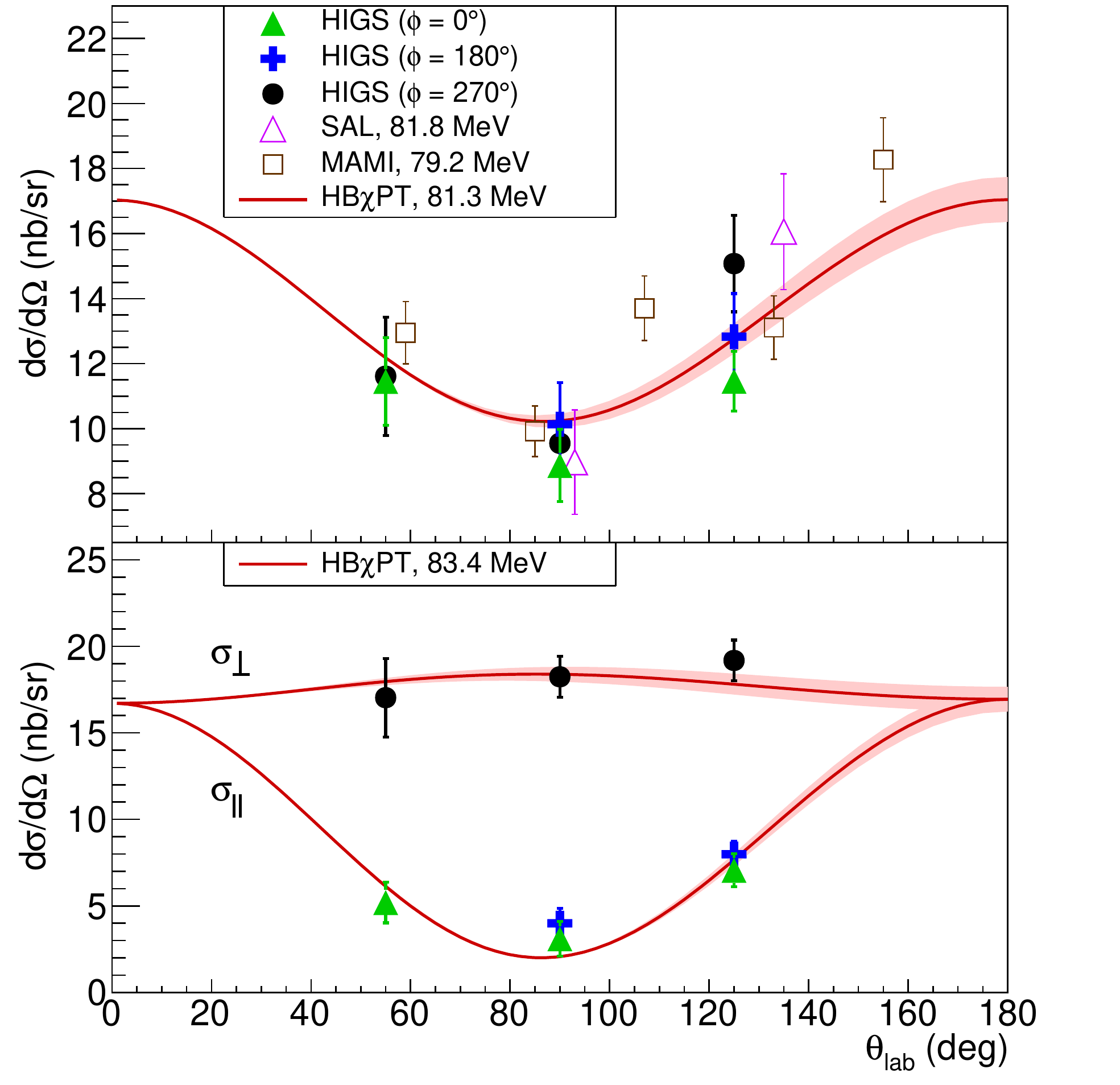}
\caption{Differential cross section results extracted from the present experiment (closed triangles, crosses, and circles) compared to the results from SAL (open triangles, note that the data point at $\theta_{\text{lab}} = 90\degree$ is plotted with a slight offset in $\theta_{\text{lab}}$ for better viewing)~\cite{MacGibbon:1995in} and MAMI (open squares)~\cite{deLeon:2001dnx}. The HIGS results in the upper (lower) panel were obtained using the circularly (linearly) polarized gamma-ray beam at 81.3\,MeV (83.4\,MeV). The error bars shown are the statistical and point-to-point systematic uncertainties added in quadrature. The curves with $1\sigma$ error bands are the theoretical cross sections implied by our measured polarizabilities using the $\chi$EFT framework.
\label{fig:DXS}}
\end{figure}

\begin{figure}[!htp]
\centering
\includegraphics[width=1\columnwidth]{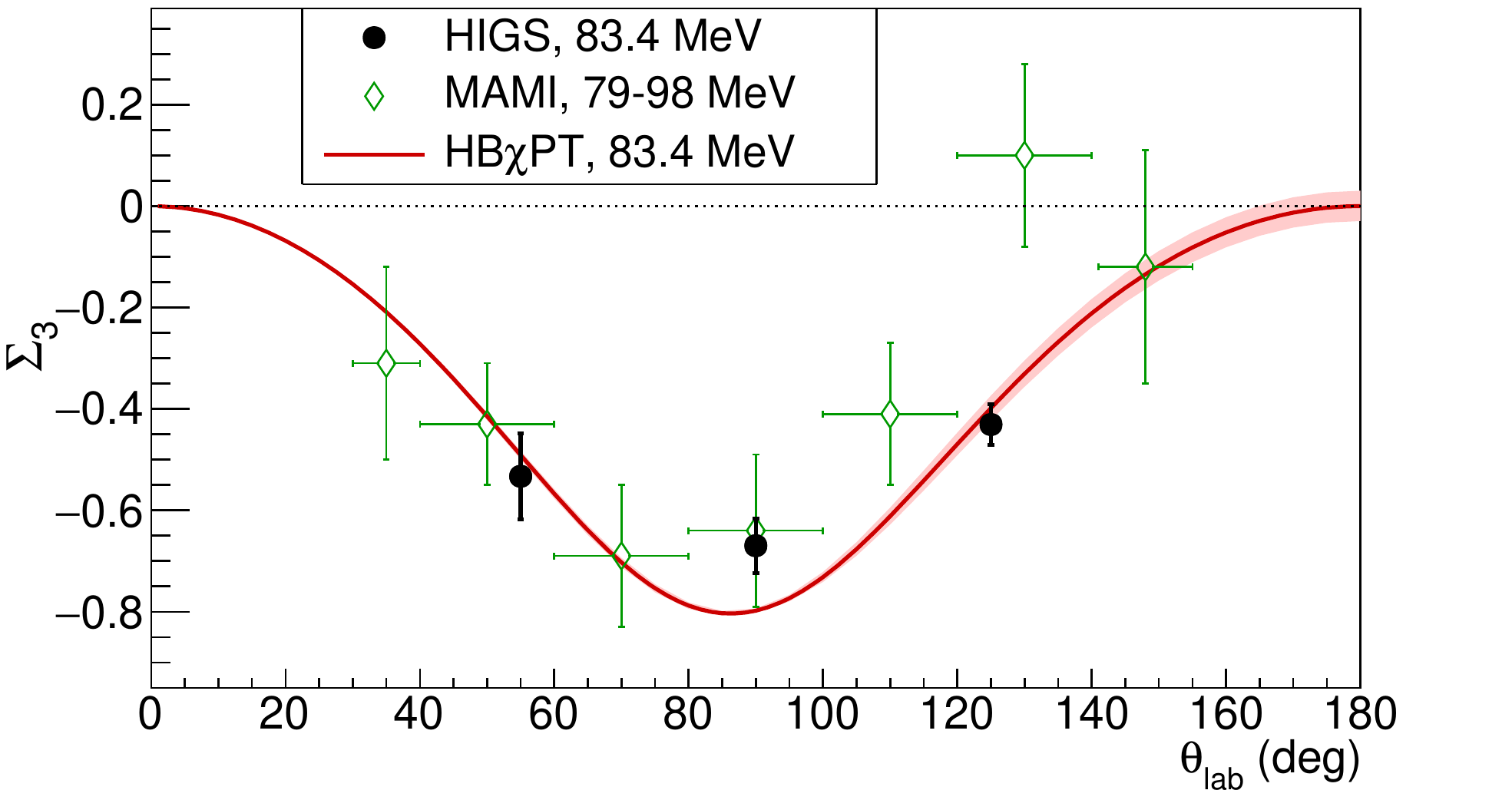}
\caption{$\Sigma_3$ obtained in the present experiment (circles) compared to the results from MAMI (diamonds)~\cite{Sokhoyan:2016yrc}. The error bars include statistical and point-to-point systematic uncertainties only. The curve with the $1\sigma$ error band is the theoretical $\Sigma_3$ implied by our measured polarizabilities using the $\chi$EFT framework.  (Note that we did not fit to $\Sigma_3$ but to the individual cross sections.)
\label{fig:BeamAsymmetry}}
\end{figure}

The extracted differential cross sections for circularly and linearly polarized photon beams are listed in Table~\ref{tab:DXS} and plotted in Fig.~\ref{fig:DXS}. Note that the Compton cross section is insensitive to the circular polarization of the photon beam for an unpolarized target, the results of the present work using the circularly polarized beam are compared to the cross sections obtained using unpolarized tagged photon beams from the Saskatchewan Accelerator Laboratory (SAL) at an incident photon energy of 81.8\,MeV~\cite{MacGibbon:1995in} and the Mainz Microtron (MAMI) facility at 79.2\,MeV~\cite{deLeon:2001dnx} in Fig.~\ref{fig:DXS}, which shows a good agreement. Given that the linear polarization direction of the incident photon beam was horizontal in the plane of $\phi = 0\degree/180\degree$, the in-plane (out-of-plane) detectors were at azimuthal angles $\phi = 0\degree \text{ or } 180\degree$ ($\phi = 270\degree$). The photon beam asymmetry $\Sigma_3$ is defined as
\begin{equation}
    \Sigma_3 = \frac{\sigma_{\parallel} - \sigma_{\perp}}{\sigma_{\parallel} + \sigma_{\perp}},
\end{equation}
where $\sigma_{\parallel}$ ($\sigma_{\perp}$) is the in-plane (out-of-plane) differential cross section. The $\Sigma_3$ values at $\theta=55\degree$, $90\degree$, and $125\degree$ were determined from the measured differential cross sections listed in Table~\ref{tab:DXS} where $\sigma_{\parallel}$ at $\theta=90\degree$ and $125\degree$ were assigned the weighted average of cross sections at $\phi = 0\degree$ and $180\degree$. In Fig.~\ref{fig:BeamAsymmetry}, the $\Sigma_3$ results from HIGS are compared to those from MAMI~\cite{Sokhoyan:2016yrc}, which shows a good agreement. 

\begin{table}[!t]
\begin{center}
\caption{Differential cross section results at $81.3$ and $83.4$\,MeV together with the statistical, point-to-point systematic, and normalization systematic uncertainties.}
\begin{tabular*}{1\columnwidth}{@{\extracolsep{\fill}} l c  c c c c}
\noalign{\smallskip}
\hline\hline\noalign{\smallskip}
$\theta_{\text{lab}}$  &$\phi$  & $d\sigma/d\Omega$ & Stat & Point to point &Normalization\\
\noalign{\smallskip}
 & &  (nb/sr) & (nb/sr) &Syst (nb/sr) &Syst (nb/sr) \\
\noalign{\smallskip}\hline\noalign{\smallskip}
\multicolumn{6}{c}{Circularly polarized gamma-ray beam at $81.3$\,MeV} \\
\noalign{\smallskip}
 $55^\circ$    & $0^\circ$    &11.45   &$\pm0.90$   &$\pm0.99$  &$\pm0.26$ \\
 $55^\circ$    & $270^\circ$  &11.61   &$\pm0.86$   &$\pm1.60$  &$\pm0.26$ \\
\noalign{\smallskip}\noalign{\smallskip}
 $90^\circ$    & $0^\circ$    &8.87    &$\pm0.73$   &$\pm0.84$  &$\pm0.20$ \\
 $90^\circ$    & $180^\circ$  &10.14   &$\pm0.76$   &$\pm1.03$  &$\pm0.23$ \\
 $90^\circ$    & $270^\circ$  &9.55    &$\pm0.63$   &$\pm0.53$  &$\pm0.21$ \\
\noalign{\smallskip}\noalign{\smallskip}
 $125^\circ$   & $0^\circ$    &11.46   &$\pm0.68$   &$\pm0.62$  &$\pm0.26$ \\
 $125^\circ$   & $180^\circ$  &12.83   &$\pm0.73$   &$\pm1.11$  &$\pm0.29$ \\
 $125^\circ$   & $270^\circ$  &15.08   &$\pm0.69$   &$\pm1.31$  &$\pm0.34$ \\
\noalign{\smallskip}\noalign{\smallskip}
\multicolumn{6}{c}{Linearly polarized gamma-ray beam at $83.4$\,MeV} \\
\noalign{\smallskip}
 $55^\circ$    & $0^\circ$    &5.19    &$\pm1.01$   &$\pm0.58$  &$\pm0.12$  \\
 $55^\circ$    & $270^\circ$  &17.03   &$\pm1.17$   &$\pm1.95$  &$\pm0.38$  \\
\noalign{\smallskip}\noalign{\smallskip}
 $90^\circ$    & $0^\circ$    &3.10    &$\pm0.94$   &$\pm0.37$  &$\pm0.07$  \\
 $90^\circ$    & $180^\circ$  &3.99    &$\pm0.83$   &$\pm0.28$  &$\pm0.09$  \\
 $90^\circ$    & $270^\circ$  &18.24   &$\pm0.93$   &$\pm0.73$  &$\pm0.41$  \\
\noalign{\smallskip}\noalign{\smallskip}
 $125^\circ$   & $0^\circ$    &7.06    &$\pm0.77$   &$\pm0.55$  &$\pm0.16$  \\
 $125^\circ$   & $180^\circ$  &7.99    &$\pm0.66$   &$\pm0.35$  &$\pm0.18$  \\
 $125^\circ$   & $270^\circ$  &19.18   &$\pm0.92$   &$\pm0.74$  &$\pm0.43$  \\
\noalign{\smallskip}\hline\hline
\end{tabular*}
\label{tab:DXS}
\end{center}
\end{table}

The cross sections obtained in this work were fitted using heavy-baryon chiral perturbation theory (HB$\chi$PT)~\cite{McGovern:2012ew} with and without the BSR constraint to extract the proton polarizabilities. The beam profiles were used to generate energy-weighted predictions for the cross sections. For the 16 HIGS data points of Fig.~\ref{fig:DXS}, the point-to-point systematic and statistical uncertainties were added in quadrature, while the normalizations of the circularly and linearly polarized beams were allowed to float independently.  However, a single-energy fit without floating normalizations gives negligibly different results.

\begin{figure}[htp]
\centering
\includegraphics[width=0.75\columnwidth]{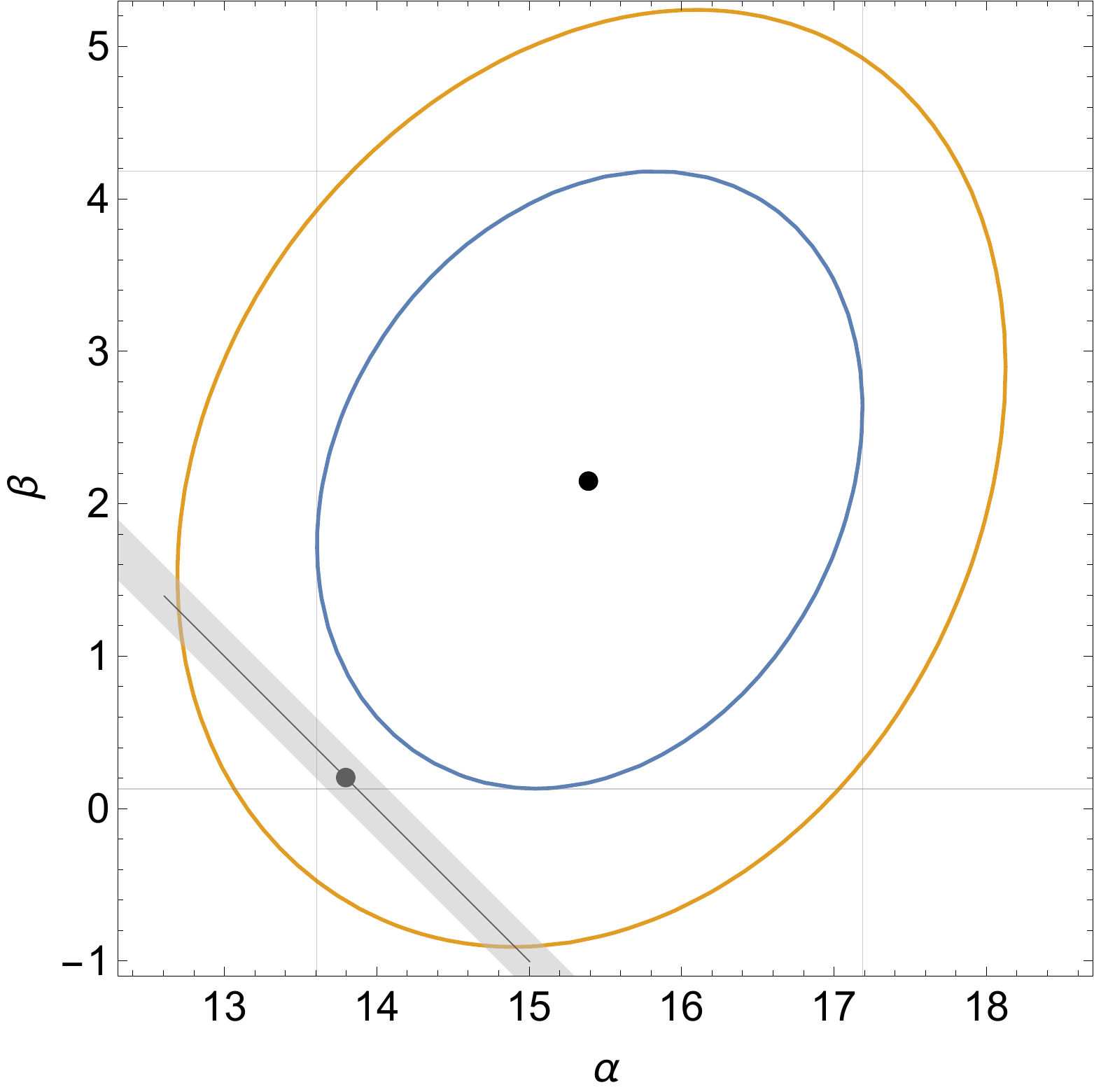}
\caption{The 1$\sigma$ ($\chi_{\text{min}}^2$+2.3, outer ellipse) and $\chi_{\text{min}}^2$+1 (inner ellipse) contours of the sum-rule-free fit of $\alpha_{E1}^p$ and $\beta_{M1}^p$ in this work. The gray band represents the BSR of Eq.~(\ref{eq:BSR}). The two dots are the central values with and without the BSR constraint.
\label{fig:Contour}}
\end{figure}

With the BSR constraint Eq.~(\ref{eq:BSR}), 16 data points were fitted with three parameters (one is $\beta$ and the other two are the normalization factors), which gives
\begin{equation}
\begin{split}
\alpha_{E1}^p &= 13.8\pm1.2_{\rm stat}\pm0.1_{\rm BSR}\pm0.3_{\rm theo},\\
\beta_{M1}^p &= 0.2\mp1.2_{\rm stat}\pm0.1_{\rm BSR}\mp0.3_{\rm theo},
\end{split}
\label{eq:withBSR}
\end{equation}
with $\chi^2=14.7$ for 13 degrees of freedom. 
The statistical errors here and in Eq.~(\ref{eq:withoutBSR}) were obtained from the $\chi_{\text{min}}^2$+1 intervals.
An estimate of the theoretical uncertainties follows the method of the convergence study in Refs.~\cite{McGovern:2012ew,Griesshammer:2015ahu}, and gives the same results. 
A fit without the BSR constraint was also performed on the 16 data points by varying both polarizabilities, which leads to
\begin{equation}
\begin{split}
\alpha_{E1}^p &= 15.4\pm1.8_{\rm stat},\\
\beta_{M1}^p &= 2.1\pm2.0_{\rm stat},
\end{split}
\label{eq:withoutBSR}
\end{equation}
with $\chi^2=13.2$ for 12 degrees of freedom. We do not quote theoretical uncertainties as this extraction is only given as a consistency check. 
Figure~\ref{fig:Contour} shows the results of the extraction of $\alpha_{E1}^p$ and $\beta_{M1}^p$ with and without the BSR constraint. We see that the one-parameter fit and its 1$\sigma$ limits are contained within the 1$\sigma$ ellipse of the two-parameter fit, which validates the use of the BSR in the former.

There is some variation between our results and the values of Eq.~(\ref{eq:EFTproton}) extracted from the same theoretical framework based on the world database of unpolarized data over a much wider energy range. This variation also pertains to the current PDG values~\cite{Zyla:2020zbs}
\begin{equation}
\alpha_{E1}^p = 11.2\pm0.4,\quad \beta_{M1}^p = 2.5\pm0.4.
\end{equation}
The pioneering extraction from the $\Sigma_3$ data obtained by MAMI~\cite{Sokhoyan:2016yrc}, while in better agreement with both of these,  has a sufficiently large uncertainty that it is also in agreement with our extraction. Fitting to cross sections rather than $\Sigma_3$ gives our result a smaller uncertainty, though conducted at a lower average energy.

Recently, there have been other determinations from the full set or a subset of the same unpolarized data, in different theoretical frameworks; for examples, see Refs.~\cite{Lensky:2014efa,Pasquini:2017ehj,Krupina:2017pgr}. At lower energies, different theories tend to agree better; therefore, high-precision data in the energy region of our experiments have an important role to play in resolving the discrepancy. The present work, as the first nanobarn-level Compton scattering measurements at HIGS, demonstrated that with the improved statistics of the HIGS data, one can extract the proton polarizabilities with better precision.

In summary, new measurements on Compton scattering from the proton were performed at HIGS below the pion-production threshold. The polarized cross sections were extracted for the first time, and unpolarized cross sections of this work are consistent with the global database. The sum-rule-free extraction of $\alpha_{E1}^p$ and $\beta_{M1}^p$ using the χEFT framework is compatible with the BSR. This work provided a novel experimental approach for Compton scattering from the proton in low energies and strongly motivates new high-precision measurements at HIGS to improve the accuracy in proton polarizabilities determinations.

We acknowledge the support of the HIGS accelerator staff for the delivery of high-quality gamma-ray beams and the help with the experimental setup. We acknowledge the contributions of Daniel Phillips to this work. This work is funded in part by the U.S. Department of Energy under Contracts No.~DE-FG02-03ER41231, No.~DE-FG02-97ER41033, No.~DE-FG02-97ER41041, No.~DE-FG02-97ER41046, No.~DE-FG02-97ER41042, No.~DE-SC0005367, No.~DE-SC0015393, No.~DE-SC0016581, and No.~DE-SC0016656, National Science Foundation Grants No.~NSF-PHY-0619183, No.~NSF-PHY-1309130, and No.~NSF-PHY-1714833, UK Science and Technology Facilities Council Grants No.~ST/L005794/1 and No.~ST/P004423/1, and funds from the Dean of the Columbian College of Arts and Sciences at The George Washington University, and its vice president for research. We acknowledge the financial support of the Natural Sciences and Engineering Research Council of Canada and the support of Eugen-Merzbacher Fellowship. 

\textit{Note added}.--Recently, new results on polarized Compton scattering from the proton reported by Mornacchi \textit{et al.} (A2 Collaboration)~\cite{Mornacchi:2021} have been posted.

\end{document}